\documentclass[twocolumn,trackchanges]{aastex7}

\usepackage{natbib}
\usepackage{xcolor}

\begin{document}

\title{MMS Insights into CME Driven Sub-Alfvénic Solar Wind at 1 AU}
\author[orcid=0000-0000-0000-0001]{Harsha Gurram}
\affiliation{Depaartment of Astronomy, University of Maryland College Park, Maryland, USA}
\affiliation{NASA Goddard Space Flight Center, Greenbelt, Maryland, USA.}
\email[show]{hgurram@umd.edu}  

\author[orcid=0000-0002-4768-189X]{Li-Jen Chen }
\affiliation{NASA Goddard Space Flight Center, Greenbelt, Maryland, USA.}
\email{li-jen.chen@nasa.gov}

\author[orcid=0000-0001-6315-1613]{Matthew R. Argall}
\affiliation{Space Science Center, University of New Hampshire, Durham, NH, USA}
\email{matthew.argall@unh.edu }

\author[orcid=0000-0003-2160-7066]{Subash Adhikari}
\affiliation{Department of Physics and Astronomy, University of Delaware, Newark, DE, USA}
\email{ }

\author[orcid=0000-0002-4313-1970]{Lynn B. Wilson}
\affiliation{NASA Goddard Space Flight Center, Greenbelt, Maryland, USA.}
\email{lynn.b.wilson@nasa.gov}

\author[orcid=0000-0001-9425-2281]{Jason R. Shuster}
\affiliation{Space Science Center, University of New Hampshire, Durham, NH, USA}
\email{Jason.Shuster@unh.edu }

\author[orcid=0000-0002-4313-1970]{Victoria D. Wilder}
\affiliation{Laboratory of Atmospheric and Space Physics, University of Colorado Boulder, Boulder, CO, USA}
\email{Victoria.Wilder@lasp.colorado.edu }

\newcommand{\scr}[1]{_{\mbox{\protect\scriptsize #1}} }

\begin{abstract}

We report the properties of electron distributions and turbulence during a Coronal Mass Ejection (CME) in April 2023 observed by Magnetospheric Multiscale (MMS). The CME exhibits a clear sheath and magnetic cloud (MC), and within the MC, the solar wind becomes sub-Alfvénic for two hours.  We investigate plasma and turbulence properties of the sub-Alfvénic CME wind and compare them with those in the super-Alfvénic solar wind in the MC and CME sheath. Electrons within the sub-Alfvénic MC show significantly higher temperatures than those in the CME sheath and the super-Alfvénic MC, with their one-dimensional distributions revealing super-thermal tail and a depletion in electron populations between 15--50~eV. Within the CME sheath, isolated regions of electron heating are observed, where parallel energy flux is enhanced up to $\sim1$keV. Magnetic field fluctuations within the sub-Alfvénic MC interval exhibit negligible cross helicity and steeper-than-Kolmogorov scaling in the inertial range, with no clear spectral break. These fluctuations also show reduced intermittency at ion and sub-ion scales, emerging intermittency at electron scales, and weak magnetic compressibility. Together, these observations point to the presence of weak magnetohydrodynamic (MHD) turbulence within the sub-Alfvénic MC, resembling conditions commonly observed in planetary magnetospheres such as Jupiter’s.
\end{abstract}



\section{Introduction} 

Coronal Mass Ejections (CMEs) are macro-scale magnetized structures containing plasma and magnetic field expelled from the Sun and are one of the key drivers of the space weather at the Earth \citep{David_howard,1993_Gosling}. CMEs can travel faster than the ambient solar wind driving an interplanetary (IP) shock ahead of them, that includes the shock front and the intervening CME sheath region. CMEs are commonly accompanied by a magnetic cloud (MC) structure with 1) enhanced magnetic field, 2) smooth rotation of the magnetic field direction and 3) depressed proton temperature and plasma beta \citep{burlaga1981magnetic,klein1982interplanetary,kilpua2017coronal}. Typically the plasma within the CME MC expands as it evolves, and when it reaches the Earth, the conditions within are super-Alfv\'{e}nic $M_A\sim5$, where $M_A$ is the Alfv\'{e}n mach number defined as the ratio of bulk ion flow to the Alfv\'{e}n speed. Under very rare circumstances, the solar wind inside MCs can become sub-Alfvénic $M_A<1$, typically during intervals of exceptionally low plasma densities and enhanced interplanetary magnetic field (IMF) strengths, both of which act to increase the Alfvén speed \citep{hajra2022subalfvenic}. A statistical survey by \citet{hajra2022subalfvenic} identified 30 near-Earth sub-Alfvénic solar wind intervals between 1973 and 2020, driven predominantly by CME-associated MCs and, in some cases, by their interaction with high-speed streams (HSSs). The occurrence of these intervals is strongly solar-cycle dependent, with 20 of the 30 events concentrated during Solar Cycle~23 (1996--2008), reflecting the sensitivity of sub-Alfvénic conditions to the prevailing level of solar activity.



\begin{figure*}[ht!]
\plotone{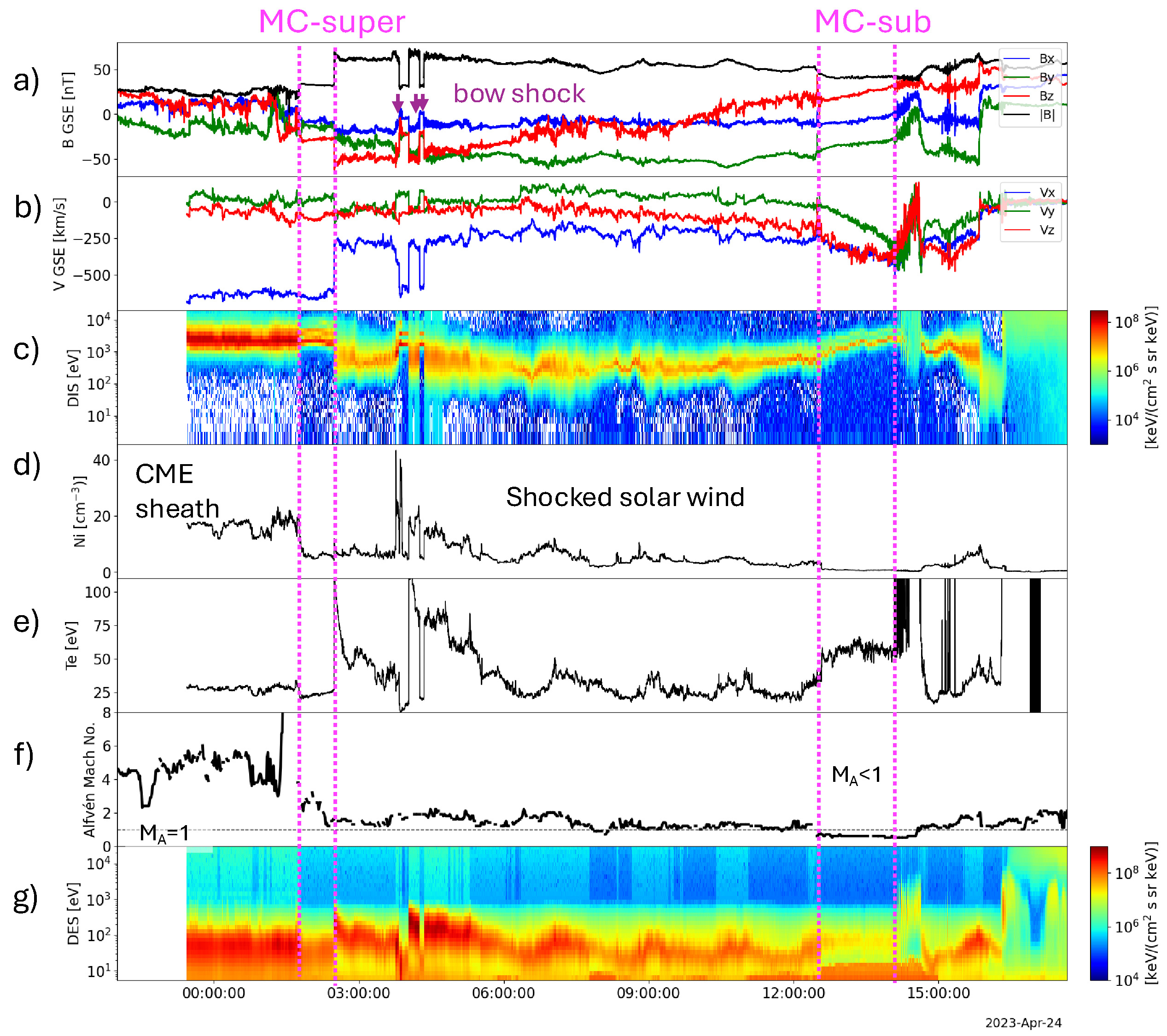}
\caption{MMS 2 observations of sheath, super-Alv\'{e}nic MC (marked MC-super) and sub-Alfv\'{e}nic MC (marked MC-sub) driven by the CME on 2023-04-23. a) Magnetic field, b) ion velocity, c) omni-directional ion energy flux, d) electron density, e) electron temperature, f) Alfv\'{e}n Mach number (from OMNI)  and g) omni-directional electron energy flux. The black dotted line in panel f) indicates $M_A=1$.}
\label{fig:event_overview}
\end{figure*}

One such rare occurrence took place on April 23$^{rd}$ 2023, a CME flux rope erupted on the surface of the Sun, driving a magnetic cloud. On April 24$^{th}$ 2023, the CME-MC passed over the Magnetospheric Multiscale \citep[MMS;][]{burch2016} spacecraft which were positioned at the dawn flank of the magnetopause ({$X_{\scr{GSE}}=[10.7, -7.9, -6.8]R_{\scr{E}}$}, in Geocentric Solar Ecliptic (GSE) coordinates) during its inbound trajectory.  Fig.~\ref{fig:event_overview} provides an overview of the April 2023 CME event, the IP shock driven by the CME arrives around 17:00 UT and MMS sees the CME sheath region from 17:40 to 01:10~UT. The figures show data from Fast Plasma Investigation \citep[FPI;][]{FPI} and Fluxgate Magnetometer \citep[FGM;][]{Russell_2016} instruments, which provided observations of plasma properties and magnetic field conditions from the MMS spacecraft. From 02:00~UT to about 12:40~UT MMS observes the shocked solar wind as marked in the figure. MMS encountered magnetic cloud in two different intervals labeled as MC-super and MC-sub in Fig.~\ref{fig:event_overview}, each characterized by different solar wind conditions. MC-super corresponds to the leading edge of the MC, where the solar wind remained super-Alfvénic, while MC-sub represents the MC’s interior, where sub-Alfvénic solar wind conditions prevailed. The magnetic fields within these two MC intervals, along with the smooth rotation in the magnetic field following MC-super, resemble the MC interval observed by the Wind spacecraft \citep{Li_jen_2024}, and are consistent with our identification of these intervals as MCs. The sub-Alfv\'{e}nic ($M_A \sim 0.6$) solar wind conditions within MC-sub was confirmed by measurements from OMNI spacecraft (shown in Fig.~\ref{fig:event_overview}e) and is caused by the low density and high interplanetary magnetic field (IMF) strength in the MC rather than a decrease in the solar wind speed \citep{Li_jen_2024}. MMS observed long-duration of sub-Alfvénic solar wind conditions for the first time, lasting approximately two hours, during which it recorded one hour of steady sub-Alfvénic flow followed by a 40-minute interval in which the typical unified structure of Earth’s magnetosphere—comprising the bow shock and magnetotail was split into Alfv\'{e}n wings \citep{Li_jen_2024}. The magnetosphere transforms into of Alfv\'{e}n wings due to the disappearance of a bow shock, a direct consequence of the sub-Alfv\'{e}nic flow \citep{Li_jen_2024} and recent work have shown the formation of Alfv\'{e}n wings, only during long-duration sub-Alfv\'enic solar wind interval ($\gtrsim 1$) using global magnetohydrodynamic (MHD) simulations and have been confirmed by the MMS observations \citep{Gurram_2024,burkholder_global_2024,Chen_yuxi}. A study by \citet{Argall2025} have studied turbulence inside the CME sheath and how turbulence evolves anistropically from Lagrange point (L1) to Earth using the Wind and MMS spacecraft. The study shows that while the inertial-range MHD spectral slope remains unchanged, the spectral break shifts to smaller spatial scales near the ion inertial length, and the sub-ion spectral slope steepens, indicating enhanced energy at these scales and a possible reverse cascade. However, turbulence within sub-Alfvénic magnetic clouds at Earth has not previously been investigated, which is one of the key objectives of this study. These studies enable the first-ever exploration of such phenomena at Earth, made possible by the unprecedented measurement capabilities by MMS. \\

In this work, we present the first detailed analysis of \textit{in situ} plasma and turbulence properties of steady sub-Alfv\'{e}nic solar wind properties at 1 AU utilizing observations from the MMS. We performed a comparative analysis between two MC intervals i.e. sub-Alfv\'{e}nic and super-Alfv\'{e}nic MCs, with particular emphasis on the plasma characteristics and turbulence within these magnetically dominated structures originating from the solar corona. Sub-Alfvénic solar wind has been observed by Parker Solar Probe in a variety of contexts, including flows associated with active-region outflows and within CME wakes (e.g., \citet{Adhikari2026Alfven,Jagarlamudi2025_ada3d3}). Parker Solar Probe has also provided sustained \textit{in situ} measurements of sub-Alfvénic solar wind close to the Sun, including repeated crossings of the Alfvén surface (e.g., \citet{Bando_2022_sub_super,Jiao_2024_Sub_Super_PSP}). More recent multi-spacecraft analyses highlight the evolving and dynamic nature of the Alfvén surface and its variability in time and space \citep{Badman2025_arxiv2509_17149}. In this work, we will discuss the insights MMS dataset offers into the plasma properties and turbulence of sub-Alfv\'{e}nic solar wind plasma at 1 AU. These observations help bridge knowledge across planetary systems, revealing how sub-Alfv\'{e}nic flows shape the plasma environments of both local and distant bodies—from Mercury and the Galilean moons to exoplanets orbiting close to their stars.



\section{Solar Wind Properties inside CME-Driven Magnetic Clouds and Sheath}
\label{sec:solarwind}
\begin{figure*}[ht!]
\includegraphics[width=1\linewidth]{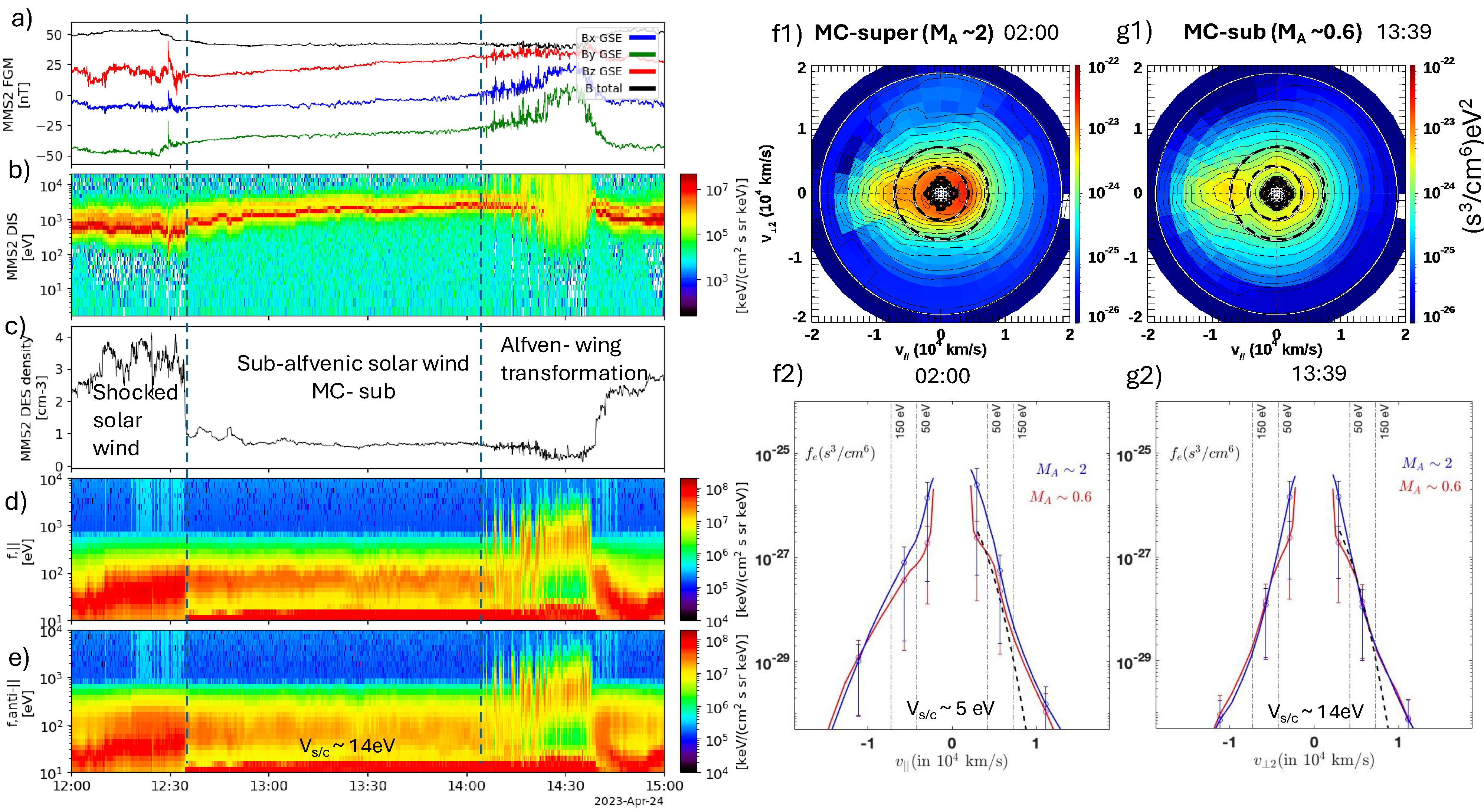}
\caption{(a) Magnetic field components, (b) omni-directional ion energy flux, (c) electron density, (d) parallel and (e) anti-parallel electron energy fluxes highlighting the sub-Alfv\'{e}nic MC (MC-sub). Panels (f1) and (g1) show electron energy velocity distribution functions (VDFs) inside the super-Alfv\'{e}nic (MC-super) and sub-Alfv\'{e}nic (MC-sub) MCs, respectively. Dashed circles in the eVDFs mark the energy levels at spacecraft potential ($V_{s/c}$), 50eV and 150eV. Panels (f2) and (g2) show 1D cuts of electron VDFs along the $_\parallel$ and $_{\perp1}$ directions respectively inside MC-super and MC-sub. Low-energy photoelectron contamination is removed from all VDFs, which leads to a lack of electron counts at velocities below $|v| \leq 2000$ km/s \citep{Gershman2019}. Error bars in panels (f2) and (g2) represent the uncertainties due to Poisson counting statistics, provided as part of the data products from FPI \citep{FPI}.}
\label{fig:sub_super_alfven}
\end{figure*}
 
In this section, we analyze the solar wind properties observed inside a CME-driven MCs, focusing on two distinct intervals. During the first MC interval, the solar wind exhibited super-Alfv\'{e}nic (MC-super) conditions, whereas during the second interval, the solar wind was sub-Alfv\'{e}nic (MC-sub) and we compare the plasma properties inside these two MCs: 1) MC-super from 01:50~UT to 02:29~UT and 2) MC-sub from 12:30~UT to 14:04~UT. The key plasma parameters for both intervals are summarized in Table~\ref{Tab:sub_super_alfven}. In MC-super, the solar wind plasma had $n_e \sim 5~\mathrm{cm}^{-3}$ and $M_A \sim 2$, and contrastingly MC-sub had a proton number density of approximately $0.5~\mathrm{cm}^{-3}$ with $M_A \sim 0.6$. The MC-super was observed prior to the interplanetary magnetic field \(B_z\) reversal, with \(B_z < 0\) throughout MC-super, while MC-sub exhibited \(B_z > 0\). Additionally, MC-sub featured a stronger magnetic field magnitude in comparison to MC-super.  The plasma beta, \(\beta\), within MC-sub was nearly an order of magnitude lower than that in MC-super, indicating a more magnetically dominated plasma. Furthermore, both electron and ion temperatures were elevated in the sub-Alfv\'{e}nic MC compared to the super-Alfv\'{e}nic MC, underscoring the distinct thermodynamic conditions associated with sub-Alfv\'{e}nic solar wind.  \\


Fig.\ref{fig:sub_super_alfven}(f1 and g1) shows the 2D electron velocity distribution functions (eVDFs) observed within the two magnetic clouds. The eVDFs are shown in the $v_{||}-v_{\bot2}$ plane, where $||$ is along the magnetic field ($B$) and $v_{\bot2}$ is in the direction of $\mathbf{v_e} \times \mathbf{B}$, where $\mathbf{v_e}$ is the electron velocity vector and $\mathbf{B}$ is magnetic field vector. These 2D distributions are energy weighted ($\propto v^4f_e$) that has been demonstrated to effectively resolve distinct magnetic topologies associated with the Alfvén wing transformation \citep{Gurram_2024}. We use these energy flux densities primarily to characterize electron populations within the two MCs and CME sheath region. Both MC-super (Fig.~\ref{fig:sub_super_alfven}f1) and MC-sub (Fig.~\ref{fig:sub_super_alfven}g1)  exhibit a narrow beam in the anti-sunward direction (\(v_{||} < 0)\) which corresponds to strahl electrons, having an energy of $272$~eV at 1~AU \citep{strahl}. However, the strahl in MC-sub is noticeably less dense than that observed in MC-super. The eVDF within MC-sub exhibits significant depletion in the electron population (within the 50 eV energy contour) relative to MC-super. Furthermore, the distribution in MC-sub appears more isotropic and Maxwellian-like in the 50–100 eV energy range, whereas MC-super displays pronounced anisotropy. The electron distributions in MC-sub also differ significantly from those in the CME sheath region. In particular, electrons within MC-sub are substantially hotter, with temperatures of approximately \(\sim60~\mathrm{eV}\), compared to those in both the CME sheath and MC-super. This elevated temperature (Fig.~\ref{fig:event_overview}e) is comparable to typical suprathermal electron energies observed at 1~AU \citep{2009JGRAStverakS,bercic2020coronal}, suggesting that the electron population within MC-sub is predominantly composed of suprathermal electrons. \\

The 1D cuts of the electron VDFs ($f_e$, not energy-weighted) inside MC-sub (Fig.~\ref{fig:sub_super_alfven}f2) reveal pronounced suprathermal tails in both MCs in the anti-$\parallel$ direction. In particular, the 1D cuts (Fig.~\ref{fig:sub_super_alfven}f2 and g2) show a clear depletion in the electron population within MC-sub in the $\sim 15$--50~eV range compared to MC-super. Differences are also evident in the strahl component between 50--284~eV: the counts (or decay rate) inside MC-super decrease more rapidly than in MC-sub, indicating a stronger superthermal electron population in MC-sub within this energy range. These observations suggest that the elevated electron temperatures observed inside the sub-Alfvénic solar wind interval (Fig.~\ref{fig:event_overview}e) are primarily due to the enhanced superthermal population which have coronal origin, along with the depletion between 15--50~eV. To quantify the anisotropy observed in the eVDFs, we overplot a Maxwellian reference curve (black dashed line) over the energy range 50--150~eV in Figs.~\ref{fig:sub_super_alfven} f2 and g2. In the $v_{\bot2}$
direction, both the MC-sub (red) and MC-super (blue) distributions closely follow the Maxwellian (as shown in Fig.~\ref{fig:sub_super_alfven} g2). However, in the parallel direction ( see Fig.~\ref{fig:sub_super_alfven} f2),  the MC-sub distribution matches the Maxwellian more closely, whereas the MC-super distribution deviates significantly, indicative of a stronger temperature anisotropy.\\


\begin{table*}
\begin{center}
\begin{tabular}{|c|c|c|c|c|c|c|c|c|c|c|c|}
\hline
 & $t_0$ & $t_f$ & $M_A$ & $B$ [nT]&$\beta$ & $V_A$ [km/s] & $V_i$ [km/s]& $T_i$ [eV]& $T_e$ [eV] & $d_i$ [km]& $d_e$ [km] \\ 
 \hline
MC-super & 01:50 & 02:29
& $2.0 \pm 0.6$
& $32.5 \pm 0.5$
& $0.30 \pm 0.04$
& $325 \pm 21$
& $650 \pm 24$
& $120 \pm 13$
& $21 \pm 1.4$
& 102 
& 2.4 \\
\hline
MC-sub & 12:40 & 14:03 
& $0.60 \pm 0.04$ 
& $41 \pm 0.6$ 
& $0.05 \pm 0.006$ 
& $950 \pm 140$ 
& $510 \pm 71$ 
& $200 \pm 69$ 
& $58 \pm 4$ 
& 243 
& 5.6 \\
 \hline
\end{tabular}
\caption{Plasma parameters of the CME magnetic cloud during super-Alfv\'enic (MC-super) and sub-Alfv\'enic (MC-sub) solar wind intervals estimated using measurements from MMS~2. MMS~2 was located at during MC-super $\sim[14,-9.5,-17]R_E$ and at $\sim[11,-8.6,-7]R_E$ MC-sub in GSE coordinates. $(t_0, t_f)$ specify the beginning and end of the time interval; $M_A$ is the Alfv\'en Mach number; $B$ is the magnetic field strength; $\beta$ is the plasma beta; $V_A$ is the Alfv\'en speed; $V_i$ is the bulk ion velocity; $T_i$ and $T_e$ are the ion and electron temperatures, respectively; and $d_i$ and $d_e$ are the ion and electron inertial lengths.}
\label{Tab:sub_super_alfven}
\end{center}
\end{table*}

\begin{figure*}[ht!]
\includegraphics[width=1.1\linewidth]{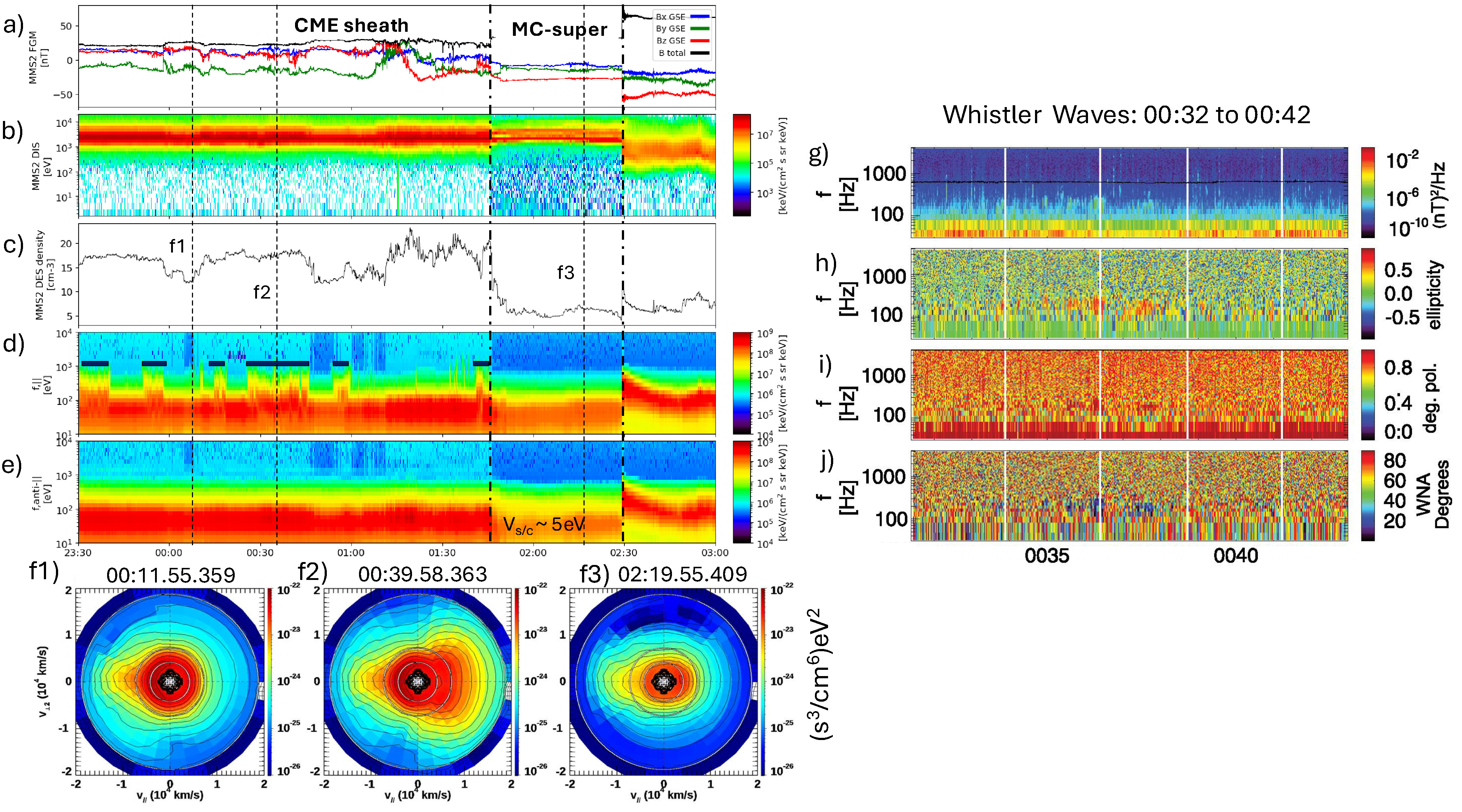}
\caption{a) Magnetic field, b) omni-directional ion energy flux, c) electron density, e) parallel and f) anti-parallel electron energy fluxes inside CME sheath and super-Alfv\'{e}nic MC. Solar wind electrons VDFs inside f1) CME sheath, f2) CME sheath with localized energetic electrons and f3) super-Alfv\'{e}nic MC.  Wave activity within the CME sheath: g) magnetic field power spectral density (PSD) with the electron cyclotron frequency ($f_{ce}$) overlaid in black; h) wave ellipticity (c) degree of polarization and (d) wave normal angle of the wave vector $\mathbf{k}$, where $0^\circ$ corresponds to field-aligned propagation and $90^\circ$ to propagation perpendicular to the background magnetic field. Polarization analysis was performed on the search-coil magnetometer data following the method of \citet{Samson1980}. }
\label{fig:sheath_MC}
\end{figure*}

MMS observed a sheath region driven by the CME from 23:30UT to $\sim$ 01:50UT (as shown in Fig ~\ref{fig:sheath_MC}). The CME sheath has a density about $\sim15$ cm$^{-3}$ with magnetic field $|B|\sim 30$nT, marked by the yellow interval in Fig.~\ref{fig:sheath_MC}. Fig.~\ref{fig:sheath_MC}d) and e) shows the parallel and anti-parallel energy flux of electrons, where parallel refers to parallel to $B$. The CME sheath exhibits electron populations with energy of $\sim 200$eV in the anti-parallel direction (Fig.~\ref{fig:sheath_MC}e), that form a narrow beam (in $v_{||}<0$) in the eVDF as shown in Fig.~\ref{fig:sheath_MC} f1). The narrow beam corresponds to strahl electrons, which have an energy of $272$eV at 1~AU \citep{strahl}. Inside the CME sheath, there are isolated regions with enhancements of parallel energy fluxes up to $\sim 1$keV highlighted by the black bar in Fig.~\ref{fig:sheath_MC}d). These isolated regions range from few seconds to about 15 minutes. Fig.~\ref{fig:sheath_MC} f2) shows the electron VDF inside these isolated energetic regions. The parallel energy flux and the distribution looks identical in the anti-parallel direction ($v_{||}<0$), however along the field aligned direction one sees a broader beam of energetic electrons extending upto $\sim20,000$km/s in the velocity space. Unlike the narrow strahl, these energetic electrons show a broader beam almost like concentric circles in $v_{||}>0$. For electrons inside MC-super and CME sheath region one major and noteworthy difference is the shape of the eVDF of the core electrons population. Inside the CME sheath the core electrons are denser, hotter ($\sim40$eV) and isotropic, contrastingly the electrons inside the MC-super are less dense, colder ($\sim20$eV) and anisotropic. The eVDF inside the MC-super exhibits strong core and strahl components, and the core population is slightly anisotropic.\\


Electrons are magnetized and therefore remain tied to the magnetic field, and as a result, electron measurements provide information about the underlying magnetic topology and field-line connectivity. Strahl electrons --- field-aligned, anti-parallel electrons at energies of $\sim$200~eV, form a strongly collimated beam and are commonly interpreted as evidence that at least one footpoint of the sampled magnetic field line is in the solar corona \citep{strahl, Gurram_2024}. In our earlier analysis of this event, energetic ($\sim1$ keV) electrons associated with the dawn and dusk Alfvén wings were found to originate from reconnection between Earth's closed field lines (CFLs) and the draped IMF, with electrons observed parallel (anti-parallel) to the field for the dusk (dawn) wing \citep{Gurram_2024}. We apply the same approach to examine the eVDFs measured inside the CME sheath. Fig.~\ref{fig:sheath_MC} f1) shows a clear strahl population inside the CME sheath, indicating that one footpoint of this field line is connected to the solar corona. The eVDF in Fig.~\ref{fig:sheath_MC} f2) similarly points to a solar coronal connection on one footpoint. Here, however, the strahl is present in the parallel direction but lacks a narrow beam, and resonant electrons are also present. We interpret the energetic electrons along these field lines as originating from CFL regions in the CME sheath, whose coronal footpoints are associated with highly energetic electron production. The absence of a narrow strahl beam, together with the resonant electron population, suggests that the field-aligned electrons may have undergone scattering by whistler waves generated below the coronal base \citep{Vocks_2003}. \\

To further investigate this, we examine the wave properties within the CME sheath using magnetic field measurements. Fig.~\ref{fig:sheath_MC} g-i) shows the magnetic field power spectral density (PSD) with the electron cyclotron frequency ($f_{ce} \approx 600 Hz$) overlaid in black, the wave ellipticity (where $+1$ denotes perfectly right-hand circular polarization), the degree of polarization (where a value of 1 indicates a fully polarized wave), and the wave normal angle ($0°$ corresponding to field-aligned propagation and $90°$ to propagation perpendicular to the magnetic field). The observed waves are centered at $\sim$200 Hz, well below $f_{ce}$, and exhibit ellipticity values near $+1$, a high degree of polarization, and wave normal angles close to $0°$, consistent with field-aligned propagation. These properties are characteristic of whistler-mode waves, which propagate preferentially along the magnetic field in the frequency range between the lower hybrid frequency and $f_{ce}$ \citep{Stix_1992}. The field-aligned propagation geometry is particularly favorable for resonant pitch-angle scattering of strahl electrons, providing a self-consistent physical picture in which coronal strahl electrons are scattered during transit, resulting in the broadened, asymmetric eVDF morphology observed in Fig.~\ref{fig:sheath_MC} f2). In addition to scattering below the coronal base, the observed beam asymmetry could also result from wave--particle interactions occurring in the solar wind or within the MC during its propagation to Earth.

\section{Turbulence properties of Sub-Alfv\'{e}nic solar wind at 1 AU} 

\begin{figure*}[ht!]
\centering
\plotone{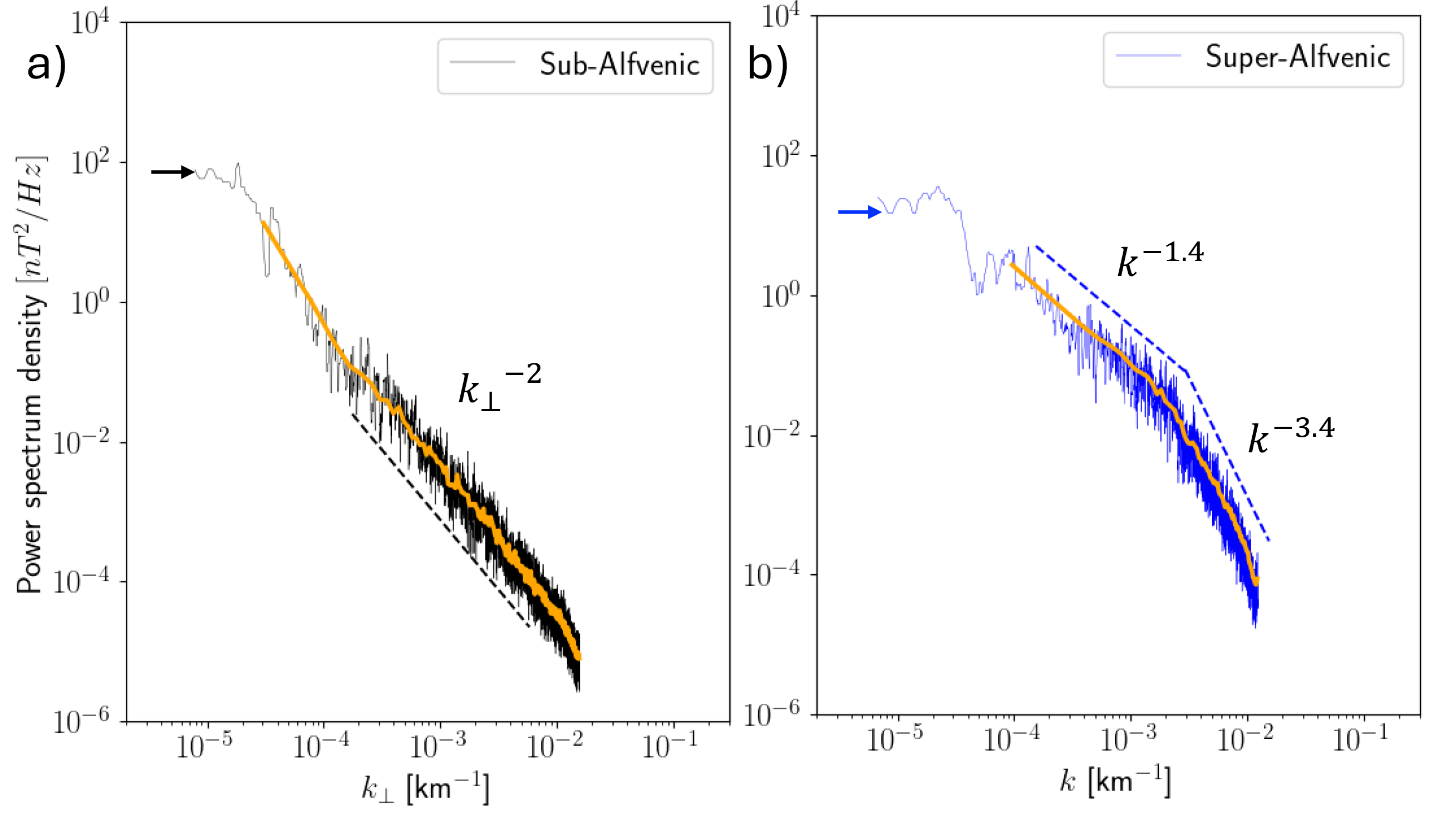}
\plotone{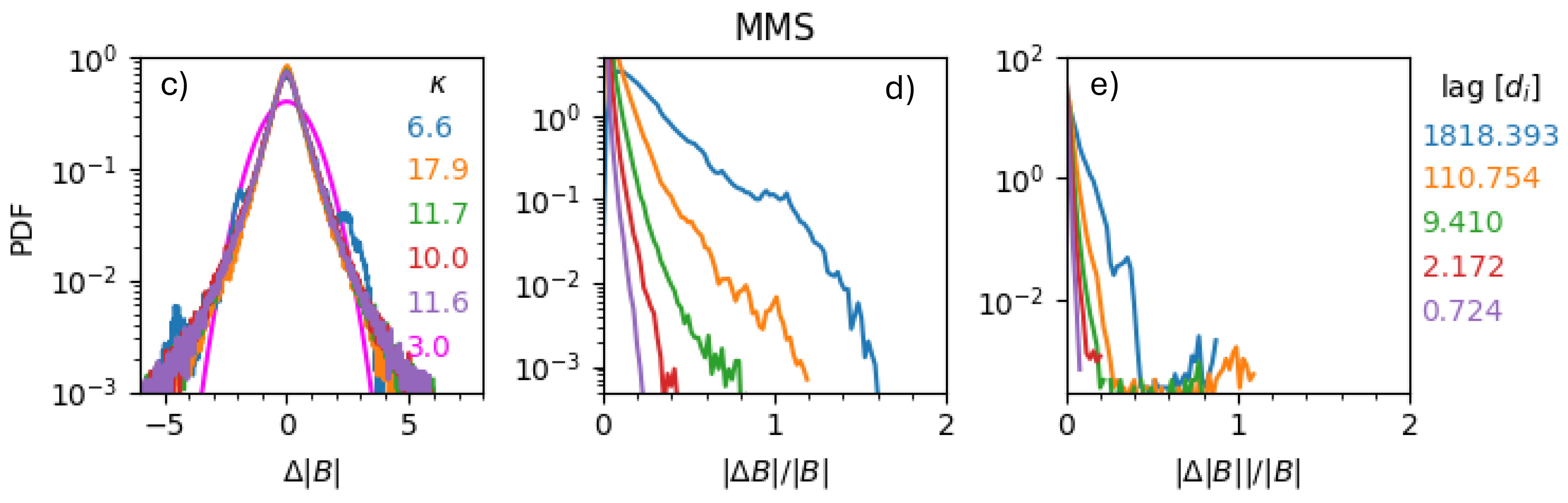}
\caption{Omnidirectional power spectrum of the magnetic field magnitude $B$ within (a) the sub-Alfv\'{e}nic (MC-sub) and (b) super-Alfv\'{e}nic (MC-super) magnetic cloud intervals, whose plasma parameters are listed in Table~\ref{Tab:sub_super_alfven}. (c) Standard normal increments of $|B|$ indicate that intermittency peaks at ion kinetic scales (orange). (d) Normalized increments demonstrate a progressive transition toward Gaussian statistics as the scale size decreases. (e) Magnetic compressibility is very small and becomes both smaller and more Gaussian with decreasing scale. (c-e) Intermittency, normalized increments and magnetic compressibility was evaluated for magnetic field fluctuations inside MC-sub. The arrows indicate the onset of the cascade, highlighting that the MC-sub exhibits an initial power spectral density approximately one order of magnitude higher than that of the MC-super.}

\label{fig:Turbulence_sub_super}
\end{figure*}

This section investigates turbulence within magnetic clouds, highlighting its comparative features relative to turbulence in CME sheaths. We employ spectral slope analysis as the key diagnostic tool to investigate the nature of fluctuations in turbulent systems across multiple scales, which captures key processes such as energy injection or driving, cascade dynamics, and kinetic dissipation \citep{Horbury_2008,Bandyppadhyay_2021,Parashar_2018}. The omnidirectional magnetic field spectrum is calculated using fast Fourier transforms (FFT) of three orthogonal components and summing the power of these components. This gives magnetic spectra as a function of the spacecraft-frame frequency and can be converted to wavenumber using Taylor Hypothesis \citep{Taylor_1938,Jokipii_1973}. For both the magnetic cloud intervals omnidirectional magnetic field spectra was computed for the low-frequency $(< 32$ Hz$ )$ utilizing FGM \citep{Russell_2016}, as shown in  Fig.\ref{fig:sub_super_alfven} a) and b). The average magnetic field and ion bulk flow vectors for the super- and sub-Alfvénic intervals, in GSE coordinates, are \(\mathbf{B}_{\scr{MC-super}} = (-8, -14, -28)~\mathrm{nT}\), \(\mathbf{V}_{\scr{MC-super}} = (-642, 22, -105)~\mathrm{km/s}\), and \(\mathbf{B}_{\scr{MC-sub}} = (-6.7, -33, 23)~\mathrm{nT}\), \(\mathbf{V}_{\scr{MC-sub}} = (-349, -154, -338)~\mathrm{km/s}\), respectively. The average angle between the magnetic field and the bulk ion flow, \(\theta_{\scr{BV}}\), is \(\sim69^\circ\) during the super-Alfv\'{e}nic and \(\sim91^\circ\) inside sub-Alfv\'{e}nic interval, indicating that fluctuations in the latter are predominantly perpendicular to the magnetic field (\(k_\perp\)).  It is important to note that for sub-Alfv\'{e}nic solar wind intervals, power spectra cannot be converted to wavenumber space using the Taylor hypothesis, \( k = f / \langle V \rangle_i \). This is in contrast to the super-Alfv\'{e}nic solar wind, where the condition \( V_{\text{SW}} (= V_i) \gg V_A \) is typically satisfied. Within the magnetic cloud MC-II, however, the solar wind speed \( V_{\text{SW}}\ll V_A \) does not strictly meet this criterion, rendering the Taylor hypothesis inapplicable \citep{Matthaeus_1982,Perri_2010}. However, one can use the assumption that the spacecraft velocity in the plasma frame is “sufficiently oblique” and connect spacecraft frequencies to the field-perpendicular wavevector $k_\bot$ i.e. $\tan\theta_{\scr{BV}}>> k_{||}/k_\bot \sim\delta u_0/V_A$, where $\delta u_0$ is the fluctuations in velocity ($\sim 33$km/s) for sub-Alfv\'{e}nic interval \citep{Perez_2021}.  The $\delta u_0$ is calculated for entire sub-Alfv\'{e}nic interval and it includes contributions from all resolved scales, rather than being restricted to the inertial range or outer-scale fluctuations alone. For the super-Alfv\'{e}nic interval we use the Taylor hypothesis and convert the frequency space into wavenumber.\\


Comparison of the magnetic fluctuations and power spectra across the MC-super, and MC-sub intervals reveals several key features: i) $\delta B/B$ for both intervals is $\sim0.01$, ii) at the injection scale, the sub-Alfv\'{e}nic interval shows magnetic fluctuation power approximately three times higher than the super-Alfv\'{e}nic case which is consistent with the enhanced magnetic field inside the MC-sub, iii) in the inertial range, the super-Alfv\'{e}nic interval displays a shallower spectral slope of $-1.4$ than the canonical \(-5/3\) whereas the sub-Alfv\'{e}nic interval exhibits a steeper slope of \(-2\), iv) the magnetic power spectrum in the sub-Alfv\'{e}nic interval does not show a clear spectral break at ion or sub-ion scales and v) the super-Alfv\'{e}nic interval exhibits a spectral break near \(\sim1~\mathrm{Hz}\), consistent with ion kinetic scales at 1~AU. A spectral slope of \(-5/3\) in the inertial range, corresponding to Kolmogorov scaling, and a slope of \(-2.8\) in the kinetic range are widely regarded as universal features of turbulence in both fast and slow solar wind streams \citep{Alexandrova_2009, Bruno_2017}.  
While the original Kolmogorov argument assumes isotropic turbulence, anisotropic magnetohydrodynamic turbulence under the framework of critical balance also predicts an inertial-range scaling of $E(k_\perp) \propto k_\perp^{-5/3}$, along the perpendicular direction \citep{Goldreich_1995}, which is broadly consistent with solar wind observations. It is worth mentioning that simulations of incompressible MHD turbulence have shown that for large scale external field, $E(k_\perp) \propto k_\perp^{-3/2}$ \citep{Boldyrev2006}. Determining whether the energy spectrum measured along the perpendicular direction follows a power exponent of -5/3 or -3/2 is still an open question \citep{podesta2009dependence}. Overall, these slopes are indicative of a well developed turbulent cascade extending all the way from fluid to kinetic scales. However, the deviations are observed inside, sub-Alfv\'{e}nic MC-sub pointing to the result of weak-MHD turbulence in the low-frequency regime. The absence of a clear spectral break and the steeper slope of $-2$ at all scales suggest a faster cascade process at all scales inside MC-sub. The magnetic spectral slope of slope $-2$ in the $k_\bot$ space with $\delta B/B$ of 0.01, and correlation lengths $l_{||}\sim1795d_i$ and $l_\bot\sim6390d_i$  gives $\tau_{A}/\tau_{nl} =(l_{||}/v_A)(\delta v/l_\bot) \sim 0.01$. The small scale fluctuations inside MC-sub with a strong background $B$ are all indicative of the presence of weak MHD-turbulence \citep{Galtier2000,Ng1996} often seen in Jovian magnetospheres \citep{Saur2002,Joachim_2021}. Previous studies \citep{podesta2010scale,bowen2018,Ervin_2025} have shown 
that the scaling of magnetic energy spectrum is dependent on both the cross 
helicity and the residual energy. They show that for more negative residual 
energy, the spectral scaling is close to $-5/3$ which falls down to $-3/2$ 
for low values of residual energy, indicating the presence of intermittent 
magnetic structures. However our studies show that even for 
larger residual energy the magnetic energy spectrum may exhibit spectral 
scaling smaller than -5/3 in the super-Alfv\'{e}nic regime. The underlying 
reason for this behavior remains unclear and left for future studies.\\

Weak MHD turbulence consists of Alfv\'{e}n waves propagating along magnetic field lines in both directions, interacting weakly by distorting one another without significantly altering their individual energies. These waves must undergo many such interactions before appreciable energy transfer occurs. In contrast, strong MHD turbulence is characterized by nonlinear interactions that lead to energy cascade on the same timescale as the wave collisions themselves leading to an imbalanced turbulence where more energy is stored in one direction \citep{LithwickGoldreich2003}.  Inside MC-sub, the nonlinear interactions inhibit the parallel cascade due to the wave resonance condition ($\omega=\pm kv_A $), which restricts energy transfer to larger perpendicular wavenumbers only. This leads to an anisotropic cascade of energy in wavevector space resulting in an imbalanced MHD turbulence \citep{Goldreich1994ApJ,Goldreich1997}. But, inside super-Alfv\'{e}nic MC, the turbulence occurs at scales immediately below the injection scale, where the turbulence cascade is hydrodynamic in nature. For scales below the Alfv\'{e}n scale, the magnetic field is amplified and dominates locally to yield weak turbulence \citep{Fornieri2021}. The resulting MHD turbulence is isotropic and the energy cascade resembles hydrodynamic turbulence, following the classic Kolmogorov scaling with exhibits a spectral break.\\

Furthermore, we examine the cross helicity as a function of scale to explore the relative alignment of magnetic and velocity field fluctuations. The normalized cross helicity is defined as $c = \frac{\langle |\mathbf{z}^{+}|^{2} - |\mathbf{z}^{-}|^{2} \rangle}{\langle |\mathbf{z}^{+}|^{2} + |\mathbf{z}^{-}|^{2} \rangle}$, where $\mathbf{z}^{\pm} = \frac{\delta \mathbf{b}}{\sqrt{4\pi n_i m_i}} \pm \delta \mathbf{v}_i$. Here $\delta \mathbf{b} = \mathbf{b}(\mathbf{x}+\mathbf{r}) - \mathbf{b}(\mathbf{x})$ is the magnetic field fluctuation, $\mathbf{r}$ is the lag, $n_i$ is the ion density, $m_i$ is the ion mass, $\delta \mathbf{v}_i$ is the ion fluid velocity fluctuation, and $\langle \cdot \rangle$ denotes a suitable average. This cross helicity can be related to the normalized residual energy $r = \frac{v^{2} - b^{2}}{v^{2} + b^{2}}$ and the alignment angle between the fluctuations via $\cos\theta = \frac{c}{\sqrt{1 - r^{2}}}$. While this relation holds both pointwise and for averages \citep{Parashar_2018}, we use an averaging window of $\sim 15$ seconds, which is about $10$ ion gyro-timescales. Overall, we find that in MC-sub the cross helicity is negligible across the inertial and kinetic scales (not shown), which may be interpreted as a lack of alignment between the velocity and magnetic field fluctuations at those scales. The smaller value of cross helicity facilitates strong nonlinear interactions \citep{Perez_Carlos_2009}, leading to rapid transfer of energy to smaller scales and enhanced dissipation within MC-sub, thereby producing a steeper slope in the resulting turbulence spectrum. However, for MC-super the angle of alignment between the fluctuations is approximately $70^\circ$ across the inertial scales (not shown), resulting in a larger cross helicity. This reduces the nonlinear interactions \citep{Smith_2009,servidio2008depression}, which in turn suppresses the energy transfer to small scales and slows down the energy cascade \citep{Smith_2009,Vasquez_2018}, resulting in a shallow energy spectrum inside MC-super.


We also investigated the intermittency of magnetic fluctuations inside the MC-sub to verify if these fluctuations corresponding to weak MHD turbulence exhibit strong intermittent structures which is typical of the solar wind as shown in several studies \citep{Burlaga_1991,Marsch_1994,Pagel_2002,Yordanova_2009}. Intermittency describes the inhomogeneity in the energy transfer between scales and is manifested as a lack of self-similarity in fluctuation distributions between scales arising from coherent structures, such as current sheets and discontinuities \citep{Kilpua_2020}. We asses the intermittency of magnetic fluctuations by computing the probability density functions (PDFs) of the increments. This involves calculating and standardizing the increments at various lags, constructing histograms to generate the PDFs, computing kurtosis, and returning the resulting distributions along with their bin centers and corresponding lag values \citep{Argall2025,Good_2020}. 
Fig.~\ref{fig:Turbulence_sub_super}c) shows the intermittency of magnetic field determined from the PDF of $(\Delta B_i^l(t)-\mu)/\sigma$, where $\mu$ and $\sigma$ are the mean and standard deviation of $\Delta B_i^l(t)$, respectively. Here, $\Delta B_i^l(t)$ are increments of the magnetic field computed for a single spacecraft as $\Delta B_i^l(t)=B(t+\tau)-B(t)$, where $l = ⟨V_i⟩ \tau$ is the spatial lag determined from temporal lag $\tau$ assuming the Taylor approximation \citep{Argall2025}.  The spatial lag values used are: l=\{0.7, 2.2, 9.4, 110.7, 1818.4\}$d_i$. At the inertial scales, the PDFs have a kurtosis $\kappa=3$, indicating a sub-Gaussian PDF. The PDFs develop intermittent tails as scale size decreases and become super-Gaussian, $\kappa>3$, exhibiting signs of intermittent turbulence at ion scales. The magnetic fluctuations inside the sub-Alfv\'{e}nic MC exhibit less intermittency compared to the turbulent CME sheath which show a kurtosis value of $\kappa\sim 5$ at ion and sub-ion scales \citep{Argall2025}. Fig.~\ref{fig:Turbulence_sub_super}e) shows the magnetic compressibility $|\Delta |B||/|B|$ is very small and becomes smaller and more Gaussian as the scale decreases for MC-sub. This further confirms that magnetic fluctuations are incompressible at all scales and predominantly Alfv\'{e}nic. Furthermore, with $\delta n/\langle n\rangle\approx0.07/0.7=0.1$, the observed density variations are very small, indicating the sub-Alfv\'{e}nic region is largely incompressible.

\section{Conclusions and Discussions}



In this paper, utilizing MMS observations, we conduct a comparative study of sub-Alfv\'enic and super-Alfv\'enic MCs, both consisting of low-beta, magnetically dominated CME plasma at $\sim$1~AU. Typically, solar wind conditions farther from the Sun are super-Alfvénic, as expansion through interplanetary space leads to a decrease in both magnetic field strength and plasma density, reducing the Alfvén speed and resulting in higher Alfvén Mach numbers. In contrast, sub-Alfv\'enic solar wind at larger heliocentric distances, particularly near Earth, is a rare occurrence. However, a unique opportunity arose during the April 2023 CME event, which enabled observations of sub-Alfv\'enic solar wind in the near-Earth environment. The sub-Alfv\'enic conditions inside the MC resulted from an enhanced IMF and typically low solar-wind plasma density, both of which increase the Alfv\'en speed and decrease $M_A$ \citep{bowers2025messenger}. This rare event provides valuable insights into the behavior and properties of the solar wind under transient, CME-induced sub-Alfv\'enic conditions.\\

We have characterized the electron VDFs inside the CME sheath, sub-and super Alfvénic MCs. The electrons inside all these three regions show an anti-field aligned beam i.e., strahl electrons. The electrons inside the CME sheath show isolated regions of heating wherein the eVDFs show a energetic $\sim1$ keV electrons along the magnetic field. We also found that electrons inside the sub-Alfvénic MC are significantly hotter than those in the CME sheath and the super-Alfvénic MC. The sub-Alfvénic solar wind is characterized by low-density, low-beta plasma ($\beta \approx 0.03$). Typically, solar wind eVDFs consist of a thermal core and superthermal halo and strahl components, with the core contributing approximately 80--90\% of the total electron density. The FPI measurements cannot accurately capture the core thermal electrons below the spacecraft potential (5.05 eV for MC-super and 14 eV for MC-sub) in the solar wind; however, they measure the other populations with an uncertainty of 5-20\% \citep{Gershman2019_uncer}. The uncertainty in the FPI measurements is primarily determined by Poisson counting statistics and therefore depends on the plasma density, with lower relative uncertainties occurring in regions of higher density. The uncertainties in both MC-sub and MC-super lie in the range of 5--20\% (as shown by error bars in Fig.~\ref{fig:sub_super_alfven}f2--g2), within which MC-super remains distinguishable from MC-sub. The electron VDFs observed within the sub-Alfvénic MC show strong superthermal electron populations and a depletion in electrons between 15--50~eV compared to the super-Alfvénic MC, resulting in elevated temperatures inside the sub-Alfvénic MC. The depletion of electron populations below 50~eV in the sub-Alfvénic MC may be responsible for the observed low-density conditions within the structure. This depletion could be linked to formation conditions in the corona, where whistler waves generated below the coronal base can resonate with electrons, preferentially enhancing the high-energy population and producing superthermal tails while leaving the core underpopulated \citep{Vocks_2003,Vocks_2008,Vocks_2021}. Although direct observational evidence for whistler wave activity at the coronal base remains elusive, this mechanism provides a plausible explanation for the observed core depletion. We further note that this coronal process is distinct from the whistler-mode waves identified \textit{in situ} within the CME sheath, which are associated with resonant electrons (shown in Fig ~\ref{fig:sheath_MC}f2). \\

We conducted the first analysis of turbulence properties inside the sub-Alfvénic MC at 1~AU and compared the properties with the turbulent CME sheath region as well as the super-Alfvénic MC. The magnetic field fluctuations inside the MC-super and MC-sub are of the order of $\sim0.01$ less than those observed in the CME sheath. The magnetic spectra inside MC-sub exhibited a power law with steeper than Kolmogorov scaling and had a spectral slope of \(-2\) in the inertial range with no clear spectral break at ion or sub-ion scales. This is different from the turbulence inside the MC-super and CME sheath, which exhibit a power-law scaling of approximately \(k^{-5/3}\) a characteristic of MHD turbulence theory, and as observed both upstream at L1 and near Earth in the inertial range\citep{Argall2025}. Other statistical studies on sheath turbulence have shown that spectral slopes for magnetic fluctuations in the majority of cases are steeper than $-5/3$.  Sheaths had on average steeper inertial-range spectral indices than the Kolmogorov (\(k^{-5/3}\)) and Kraichnan ($k^{-3/2}$). The spectral indices span a wide range $(\sim[-2.2, -1.3])$ and the spread was larger for slow sheaths and sheaths preceded by slow solar wind \citep{Kilpua_2020,Kilpua_2021}. The CME-driven sheath observed in this event had a higher propagation and was accompanied by a slow solar wind with spectral slopes in the kinetic range of $-2.8$ at L1 and $-3.6$ near Earth, as shown by \citet{Argall2025}. \\

Recent studies using Parker Solar Probe (PSP) have, for the first time, provided detailed observations of sub-Alfvénic solar wind originating from the solar corona, below the Alfvén surface \citep{Bando_2022_sub_super,Alberti_2022,Zhang_2022,Zhao_2022,Zank_2022}. \citet{Zank_2022} analyzed the spectral characteristics of this regime and reported strong consistency with predictions from nearly incompressible MHD theory. \citet{Alberti_2022} also investigated the scaling properties of both sub- and super-Alfvénic solar wind and identified convex scaling exponents in the inertial range and linear scaling at sub-ion scales for both regimes. In contrast, the sub-Alfvénic interval at 1~AU shows a steeper magnetic field power-law slope of \(-2\) in the inertial range, suggesting a weak-MHD turbulence cascade. This is similar to the weak kinetic Alfvén wave turbulence observed at sub-ion scales in Io's flux tube \citep{Janser_2022}. The fluctuations inside MC-sub have negligible cross-helicity, resulting in stronger non-linear interactions and faster energy cascade compared to MC-super and the CME sheath. The magnetic field structures inside the MC-sub show less intermittency compared to CME sheath at the ion scales, but the fluctuations develop intermittency at kinetic scales, consistent with dissipation and current sheet formation. The density and magnetic variations are also minimal, suggesting the steady sub-Alfvénic solar wind at 1~AU is incompressible.

\begin{acknowledgments}
We gratefully acknowledge the MMS mission team and instrument principal investigators for providing access to high-quality data and for their continued support. We also thank the FPI Visualizer team for their assistance and for making their visualization tools openly accessible at https://fpi.gsfc.nasa.gov/. Our appreciation extends to the SPEDAS/pySPEDAS development team for offering robust data analysis tools that supported this study. HG acknowledges partial support of NSF grant AGS2247718.

\end{acknowledgments}




\bibliography{main}{}
\bibliographystyle{aasjournalv7}



\end{document}